\documentclass[12pt,preprint]{aastex}

\makeatletter
\def\ale{\mathrel{\mathpalette\gl@align<}}
\def\age{\mathrel{\mathpalette\gl@align>}}
\def\gl@align#1#2{\lower.6ex\vbox{\baselineskip\z@skip\lineskip\z@
\ialign{$\m@th#1\hfil##\hfil$\crcr#2\crcr\sim\crcr}}}
\makeatletter
\newcommand{\twodust}{%
\raisebox{0.3ex}{{\sc 2-D}}%
$\!\!\!\!\!\;$%
\raisebox{-0.3ex}{{\sc Dust}}}

\begin{document}

\shorttitle{Two Subclasses of Proto-Planetary Nebulae} 
\shortauthors{Meixner, Ueta, Bobrowsky, \& Speck}
\title{Two Subclasses of Proto-Planetary Nebulae:  Model Calculations}

\author{Margaret Meixner and Toshiya Ueta }
\affil{Department of Astronomy, MC-221, 
University of Illinois at Urbana-Champaign, 
Urbana, IL  61801,
meixner@astro.uiuc.edu,
ueta@astro.uiuc.edu}

\author{Matthew Bobrowsky}
\affil{ Challenger Center for Space Science Education, 
1250 North Pitt Street,
Alexandria, VA 22314,
mbobrowsky@challenger.org}

\author{Angela Speck}
\affil{Department of Astronomy, MC-221, 
University of Illinois at Urbana-Champaign, 
Urbana, IL  61801,
akspeck@astro.uiuc.edu}

\begin{abstract}

We use detailed radiative transfer models  to investigate the
differences between  the Star-Obvious Low-level-Elongated 
proto-planetary nebulae (SOLE PPNs) and
 DUst-Prominent Longitudinally-EXtended 
proto-planetary nebulae (DUPLEX PPNs) which are two subclasses
of PPNs suggested by \citet{ueta00}.   
We select one SOLE PPN, HD~161796, and one DUPLEX PPN, IRAS~17150$-$3224,
both of which are well studied and representative of their PPN classes.
Using an axisymmetric dust shell radiative transfer code, we model
these two sources in detail and constrain their mass-loss histories,
inclination angles and dust composition.
The  physical
parameters derived for HD~161796 and IRAS~17150$-$3224 demonstrate that they
are physically quite  different  
and that their observed differences  cannot be attributed to
inclination angle effects.  Both HD~161796 and IRAS~17150$-$3224 are viewed 
nearly edge-on.
However, the more intensive axisymmetric superwind mass loss  experienced by 
IRAS~17150$-$3224 (8.5$\times10^{-3}$ M$_\odot$ yr$^{-1}$
 and an 
$\rm \dot M_{equator} / \dot M_{pole} = $ 160) has created a high optical 
depth dust torus (A$_V$=37) which
obscures its central star. In contrast,
HD~161796, which underwent a lower rate superwind 
(\.M$= 1.2\times 10^{-4}$ M$_\odot$ yr$^{-1}$ and 
an $\rm \dot M_{equator} / \dot M_{pole} = $ 9), has an optically thinner 
dust shell which allows the penetration of direct star light.
Based on our analysis of the dust composition, which is constrained by
dust optical constants derived from
laboratory measurements, both objects contain oxygen rich dust, mainly
amorphous silicates, but with some significant differences.
IRAS~17150$-$3224 contains only amorphous silicates with sizes
ranging from 0.001 $\mu$m to  larger than $\sim$200 $\mu$m.  HD~161796 contains 
amorphous silicates, crystalline silicates (enstatite and forsterite),
and crystalline water ice with sizes ranging from 0.2 $\mu$m to larger
than $\sim$10 $\mu$m.
If these calculations reflect a more general truth
about SOLE vs. DUPLEX PPNs, then these two subclasses
of PPNs are physically distinct with the SOLE PPNs derived from low
mass progenitors and DUPLEX PPNs derived from high mass progenitors.

\end{abstract}

\keywords{stars: AGB and post-AGB --- stars: mass loss  
--- planetary nebulae: general --- reflection nebulae --- individual:
HD~161796, IRAS~17150$-$3224}

\section{Introduction}

The expulsion of a star's  outer envelope heralds the end
for most stars in our galaxy.  For stars like our Sun, the observed rates
at which stars lose their outer envelopes  can far exceed the interior 
nuclear burning rates, causing these stars to wither into white 
dwarfs. What physical process initiates this dramatic loss of mass?
Some clues to the mass loss mechanism 
can come from the morphology of the ejected material. But here lies
an additional puzzle.  Planetary nebulae (PNs), which are the stellar
ejecta illuminated by the stellar cores that are rapidly
evolving into white dwarfs \citep{iben95},  appear to 
be mostly axially symmetric with subclasses of bipolar and
elliptical morphologies \citep{balick87,zuckerman86}.  
On the other hand, the ejected circumstellar material of 
most asymptotic giant branch (AGB) stars, which are the precursors to 
PNs,
appear spherically symmetric as the star sheds its envelope 
\citep{neri98,habing93}.
The morphologies of planetary nebulae are predominantly shaped by the AGB
mass loss and by interaction of that mass loss with 
a fast wind from the evolving central star.  The axisymmetry of 
PNs may arise when the fast wind  expands a hot gas bubble
into previously ejected material that has an equatorial density 
enhancement \citep{frank94}.
Preferential increases of density at the equator may be created by 
interactions of the mass loss with a binary companion 
\citep{soker98,mastrodemos98} or 
with stellar rotation and a  magnetic field \citep{garcia99,blackman01}.   
Proto-planetary nebulae (PPNs), which  are objects in transition from 
AGB   stars to PNs \citep{kwok93}, 
offer  pristine fossil records of
the mass loss histories because they predate the fast wind shaping.

Observational studies of PPN morphologies have shown
that they are inherently axisymmetric \citep{meixner99,ueta00,
trammell94,hrivnak91,su00} demonstrating that the axisymmetric
structures  found in PNs predate the PPN phase.  
Axisymmetric PN morphologies are subdivided into
bipolar, elliptical  and other categories. 
Based on HST imaging study of 27 PPNs,
\cite{ueta00} suggest there are two subclasses of  PPN, 
Star-Obvious Low-level-Elongated (SOLE) and
DUst-Prominent Longitudinally-EXtended (DUPLEX), that may be
precursors to the elliptical and bipolar PNs, respectively.
SOLE PPNs have prominent central stars surrounded by low surface
brightness nebulae with single ellipses or sometimes multi-lobed structures.
DUPLEX PPNs have spectacular bipolar reflection nebulae with pinched
waists and  no direct view or a heavily obscured view of the central star.
SOLE and DUPLEX PPNs also appear
to differ in three other observational categories. Firstly, for SOLE PPNs, 
mid-infrared  (mid-IR) images of the thermal dust emission reveal  a toroidal 
structure embedded
in a larger elliptical envelope.  For DUPLEX PPNs, the mid-IR emission
appears as a bright  unresolved
core surrounded by a low surface brightness emission \citep{meixner99}.   
Secondly, the spectral energy
distributions (SEDs) for SOLE PPNs have roughly equal contributions 
from the optical and far-infrared.
On the other hand, in the SEDs for DUPLEX nebulae,
the far-infrared contribution dominates over the optical \citep{ueta00,
veen89}.
Thirdly,  the galactic scale height of the SOLE  PPNs is larger
than that of the DUPLEX PPNs which appear to be more confined to the galactic
plane \citep{ueta00}.  \citet{corradi95} have shown a similar difference
for elliptical PN versus bipolar PN. These differences are summarized
in Table \ref{tabsoledup}.

In this work, the differences between
SOLE and DUPLEX PPNs are investigated 
using an axisymmetric radiative transfer code to model
the dust shells of two  PPNs. 
Section 2 outlines the reasoning behind our choice of objects.
In section 3,  we review the model calculations and
results.  In section 4, we  compare and discuss  the derived physical
parameters of the two sources. Section 5 summarizes the conclusions.

\section{Selection of Objects}

In order to solidly substantiate that SOLE and DUPLEX PPNs are physically
distinct PPNs, as opposed to, e.g.,  the same PPN viewed at different
inclination angles, detailed model calculations of the
entire sample of PPN would be needed.  However, the computation time
to accomplish such a task is substantial.  The radiative
transfer code that we use  takes a few to a few dozen hours to compute one
model run and approximately  200 models are run in order to find a best fit.
Thus, we have instead pursued a more exploratory approach of modeling
one SOLE PPN and one DUPLEX PPN.  In order to have the best constrained
model calculations, we have
selected two well-studied PPN from the HST imaging sample:  
HD~161796, a SOLE PPN,  and IRAS~17150$-$3224 (hereafter IRAS~17150), 
a DUPLEX PPN.  These two PPNs share a number of properties of other PPNs
in their class (\S 1, \citealt{ueta00}), and therefore it is reasonable
to expect the results for these objects will be representative of their
SOLE and DUPLEX classes.  These two PPN also have some similarities.
Both objects are oxygen-rich and appear to contain silicate dust grains 
\citep{justtanont92,kwok95}. Both HD~161796 and IRAS~17150 appear to
be nearly edge-on from their optical and mid-IR 
images \citep{ueta00,meixner99}.
However, their similarities end there.  HD~161796, which is
also known as IRAS~17436+5003, has a galactic latitude of 30.9\arcdeg~
and has most likely evolved from a thick-disk  star \citep{luck90}.
IRAS~17150, on the other hand, has a galactic latitude of 3.0\arcdeg~
and is thus most certainly a population I object.  

The observed properties for these two PPNs are listed in Table \ref{tabobsprop}.
The effective temperatures of the central
stars (T$_*$) are based on spectral types of these stars: F3~Ib for
HD~161796 \citep{fernie84}  and G2~Ia for IRAS~17150. 
\citep{hu93}.  \citet{skinner94} derived a  distance to HD~161796 of 
$\sim$1 kpc which is consistent with the Hipparchos
results.  The distance to IRAS~17150 is
not as well known.   IRAS~17150  is located in the plane
of our galaxy with a galactic latititude and longitude of 3\arcdeg, 
354\arcdeg~ and, thus, we can assign a kinematic distance.  Using  its
systemic velocity of $\sim$15 km s$^{-1}$ \citep{hu93}, we find a distance of
3.6 kpc using the rotation curve information derived from HI observations
\citep{burton88}.  This kinematic distance seems reasonable in light of
the resemblance that IRAS~17150 bears to AFGL~2688 \citep{kwok98},
which has an estimated distance
of $\sim$1.2 kpc \citep{skinner97}. If we assume that IRAS~17150 and AFGL 2688 
have comparable luminosities, then we estimate a distance of 3.6 kpc for 
IRAS~17150 by comparing their bolometric fluxes.  
Extinction due to interstellar dust was estimated from the work
of \citet{neckel80}.  The expansion velocities, $v_{\rm exp}$,
are based on CO observations.  We note, however,
that for IRAS~17150, the CO observations
show a larger expansion velocity of 15 km s$^{-1}$ compared to
the 6 km s$^{-1}$ of the 1665 MHz OH masers and thus the actual
expansion velocity is uncertain due to velocity gradients in this source  
\citep{hu93} but the CO observations will be adequate for our comparison
of the two objects.

\section{Axisymmetric Models}  

For our modeling,
we use a dust radiative transfer  code, called \twodust.
The computation method is based on the iterative
scheme devised by \citet{collison91} and follows the principle of
long characteristic.  It will be discussed in detail
by \citet{ueta02}. In these calculations, the central star illuminates
an axisymmetric circumstellar dust shell, and the dust shell reddens and
scatters the starlight.  Self-consistency is achieved through
requiring global luminosity constancy at each radial grid point in the
dust shell. The density function of the dust shell is of
central interest to this work because it is directly related to the
mass loss history of the object.   Here we have adopted
a density function that can have
a toroidal interior, an elliptical (prolate or oblate) mid-region,
and a spherical outer shell.  This function embodies the idea that
mass-loss rate was spherically symmetric on the AGB (\.M$_{\rm AGB}$) 
and became axisymmetric during the superwind phase (\.M$_{\rm SW}$). 
The density functions used in our calculations are shown in 
Figure \ref{densfunct} and  have the following form:

 \begin{eqnarray}
 \rho (R, \theta)  & = & \rho_{\rm min}
  \left( R/R_{\rm min} \right)^{%
  -B \left[ 1 + C \sin ^{F} \theta
      \left(
       e^{-\left( R/R_{\rm sw} \right)^{D}} / 
       e^{-\left( R_{\rm min}/R_{\rm sw} \right)^{D}}
      \right)  
    \right]} \nonumber \\
 & & \times
  \left[ 1 + A (1 - \cos \theta)^{F}
      \left(
       e^{-\left( R/R_{\rm sw} \right)^{E}} / 
       e^{-\left( R_{\rm min}/R_{\rm sw} \right)^{E}}
      \right)  
 \right] ~~~R_{\rm min} < R < R_{\rm max}
\end{eqnarray}

\noindent where $\rho(R,\theta)$ is the dust grain mass density at radius
$R$ and latitude $\theta$, $\rho_{\rm min}$ the dust grain
mass density on the polar axis at the inner edge of the
envelope, $ R_{\rm min}$ is the inner radius of the shell,
$ R_{\rm max}$ is the outer radius of the shell,
$ R_{\rm sw}$  is the boundary between the spherical
AGB wind and the axisymmetric superwind. The first term, 
$ R/R_{\rm min} ^{-B}$, defines the radial profile of the spherical
AGB wind. The  $[1 + C...]$ term which follows $B$ in the exponent, 
defines the elliptical
mid-shell.  The second term, $[1+A...]$, defines the equatorial enhancement
of the superwind.
The user defined constants, $A$ through $F$, define the density profile.
The degree of equatorial enhancement is set by $A$, where
$1+A$ is the equator-to-pole density ratio at $R_{\rm min}$,  
and $C$, where the density drop with radius along the equator is steepened
by a factor of $(1+C)$.
The inner torus can be made torus-like or disk-like with $F$, 
while $D$ and $E$ define the mid-shell region oblate or prolate.

Figure \ref{densfunct} also shows the 2-D density functions used
in our ``best fit'' models of HD~161796 and IRAS~17150.  The user 
defined constants for the model density functions
are listed in Table \ref{tabdensfunct}.
The $\rho_{\rm min}$ is determined by the user specified optical
depth at the equator of the dust shell at a specific wavelength
and is listed in Table \ref{tabdensfunct} as
$\tau_{\rm 9.8\mu m}$ at equator.
  
The properties of the dust grains are also constrained by our modeling.
For the grain size distribution, we assume a power law plus exponential
fall off 

 \begin{eqnarray}
a^{-3.5}e^{-a/ a_0}~~~a_{\rm min} < a < \infty
\end{eqnarray}

\noindent for which a minimum grain size, $a_{\rm min}$,  and an effective
maximum grain size, $a_0$, are specified,
and for which the number of grains larger than $a_0$ is diminishingly
small \citep{kim94}. This grain size distribution was derived for
the ISM and may or may not be appropriate for circumstellar grains.
A theoretical study by \cite{dominik89} suggests that the grain size
distribution created in the circumstellar environments of AGB stars
has a steeper power law ($a^{-5}$), however, this parameter has yet
to be studied observationally. The absorption and scattering cross sections 
are calculated for each wavelength using Mie theory and dust optical
constants derived from laboratory measurements. For the radiative transfer,
we consider a fiducial grain which has  size-and-composition-averaged cross sections at each wavelength \citep{ueta01b}.  
Table \ref{tabdustprop} lists the dust properties for each source.
In a detailed study of HD~161796's mineralogy,
\citet{hoogzaad2001}  have determined that amorphous silicates 
(Mg$_{0.5}$Fe$_{0.5}$SiO$_4$, olivine)
\citep{dorschner95},
water ice \citep{warren84,bertie69}, crystalline forsterite
\citep{scott96,servoin73} and crystalline enstatite 
\citep{scott96,jaeger98} are
all present in the circumstellar dust shell.    Thus, we include all
four dust species in our model.  For IRAS~17150, however,  we needed only
amorphous silicates (Mg$_{0.5}$Fe$_{0.5}$SiO$_4$, olivine) 
\citep{dorschner95}.

We approached the modeling by starting simply and 
increasing complexity as demanded by the observations. 
The SEDs are fitted first and these
constrain the dust opacity ($\rm \tau_{9.8\mu m}$), dust grain composition,
luminosity ($\rm L_*$), stellar and dust
temperatures ($\rm T_*, T_{dust}$)
and inner radius of the dust shell ($ R_{\rm min}$). The optical
and mid-infrared images constrain $R_{\rm min}$,  $\tau_{\rm 9.8\mu m}$, 
the inclination angle ({\it i}), density function parameters $A$--$F$.
Approximately 300 models were run for IRAS~17150 and 150 models
were run for HD~161796 before convergence on a ``best fit'' model. 
We have used earlier versions of this code  to model several
objects \citep{meixner97,skinner97,ueta01a,ueta01b}.
We refer  the reader to  \citet{ueta01a}, \citet{skinner97}
and appendix A of \citet{meixner97} 
for further details on the modeling procedure,
and to \citet{ueta01b} for details on the effects of a grain-size 
distribution.  

In the following sections, we compare the
model SED with spectra and photometry  and model images with
data images for our two sources,  HD~161796 and IRAS~17150.
The photometry, spectroscopy and image data are taken from the literature.
The derived stellar and dust shell parameters
from the models are listed in Table \ref{tabobsprop}.  The
model images have the same pixel scale as the observed
images. The model mid-IR images are convolved
with a Gaussian  point spread function (PSF)  
that has a full-width half-maximum (FWHM) 
equivalent to the observed FWHM of the mid-IR images.  
The model optical images did not require smoothing for comparison
with the HST images.

\subsection{ HD~161796}

Previous model calculations for HD~161796 have assumed spherical symmetry
\citep{hrivnak89,skinner94,hoogzaad2001} and have concentrated
on fits to the SED, while here we present the first
axisymmetric model for this source and make detailed comparisons
with images and the SED.  We compare the data and model results for 
HD~161796 with  images (Fig. \ref{hd161796image})  and 
SED (Fig. \ref{hd161796sed}).  The best fit model has an
inclination angle of 90\arcdeg~ (see Tables \ref{tabobsprop},
\ref{tabdensfunct},  and \ref{tabdustprop} for other
model parameters). 
Looking at the images of HD~161796, one can see how the spectral
energy distribution is formed.  The bright central star outshines
the low level elliptical reflection nebula 
in the optical (Fig. \ref{hd161796image}). 
Both the central
star and reflection nebulosity contribute significantly 
to the  prominent  optical peak in the SED
(Fig. \ref{hd161796sed}). The contours  of 12.5 $\mu$m emission show
the distribution of thermal
radiation from the dust grains and reveal a round dust
nebula with two limb brightened peaks located on either side of the 
central star  indicative of an edge-on, optically thin
dust torus (Fig. \ref{hd161796image};
\citealt{skinner94}).  An image  
at 8.5\micron~from \citet{skinner94} 
shows that the central star is located at the center
of the dust nebula and is used for registration of the V-band
and 12.5 \micron~images.  The dust shell is responsible for
the far-infrared peak in the SED (Fig. \ref{hd161796sed}), and the mid-IR 
emission shown in Fig. \ref{hd161796image}
arises from the inner edge of this dust shell.

The model images compare reasonably well with the data images
 (Fig. \ref{hd161796image}).  In particular, the star is  visible amongst the
reflection nebulosity, even in this edge-on configuration.
The model mid-IR emission nebula has a separation of
the two limb-brightened peaks that match the data exactly.  
Indeed,  the separation of the two peaks provides a tight
constraint on $ R_{\rm min}$.  The model optical reflection
nebula  has a comparable relative brightness to the central
star as shown by the HST images.  The model image reproduces the
elliptical shape of the observed reflection nebula,  but extends
to a larger radius than observed and appears more rounded in the outer
radii.   While HST's WFPC2 PSF halo
could hide a faint spherical halo around the
elliptical reflection nebula, the model images are not a perfect match
to the data.  In an optically thin nebula, the mid-IR traces the location 
of the warm dust and  the
optical reflection nebulosity traces the location of the dust near
the central star where densities and optical light intensities are highest.  
Thus some of the slight differences between the model and data images
suggest some slight differences between the  model and the actual
dust density distribution. For example,  the two peaks in the mid-IR
nebula are rotated approximately 25\arcdeg~ compare to the model
and the eastern peak is slightly brighter than the western peak
whereas the model shows both peaks at an equal brightness.
This uneveness suggests a slight asymmetry in the axisymmetric dust
density distribution or perhaps that the central star is off center
and heating the eastern portion to a hotter temperature.  Secondly, the
mid-IR nebula is more extended than the model nebula which may be in part 
due to the PSF being non-Gaussian.  However, a larger mid-IR nebula may
be produced  if the density drop-off is slower than our model assumes.

Drastically changing the inclination angle to our
line of sight only worsens the fit to the images.  Figure
\ref{hd161796inclang} shows model images for our best fit model
at inclination angles from 0 to 90\arcdeg. As our view of the nebula
changes from edge-on (90\arcdeg) to pole-on (0\arcdeg), the optical
reflection nebula changes from an elliptical shape to a circular shape
and the two limb-brightened peaks in the mid-IR emission nebula 
become broader and fade into a circular ring.  While a 90\arcdeg~
inclination angle fits the SED and images best,  we find that 
an inclination angle as low as 60\arcdeg~ would provide
an adequate fit to the data images. Thus, we include an error bar of
$\pm$30\arcdeg.

The model SED  is in good agreement with the photometry
and ISO spectrum of this source (Fig. \ref{hd161796sed}) and the SED fit is
better than, e.g.,  that of \citet{skinner94}.
The ISO spectrum \citep{molster00}  and the model calculations
of \citet{hoogzaad2001} have 
much better spectral resolution than our model. Consequentially,
our model shows only the gross dust features such as the broad
amorphous silicate bands at 10 and 18 $\mu$m and the water ice
feature at 43$\mu$m.  The crystalline silicate features
due to forsterite  and enstatite (24 and 34$\mu$m), which
are reproduced by  \citet{hoogzaad2001}, are not reproduced
by our model due to the coarse sampling of wavelengths.
The near-IR radiation is slightly over-estimated by our model
because the model water ice absorption features in the near-infrared
(3, 4.6, and 6 $\mu$m) are too shallow.
Our assumed mass fraction of water ice, 2.5\%, is substantially less than
that of \citet{hoogzaad2001} who found 24\%.  In the model of
\citet{hoogzaad2001}, most of the water ice exists as a mantle
of crystalline silicate coated dust grains and this dust model causes
the mass fraction of water ice to be quite high. In our model calculations,
we found that the 43$\mu$m water ice feature was too pronounced with a 
24\% mass fraction. However, the near-IR flux levels were better fit 
with a  24\% mass fraction of water ice.  The mid-IR flux, (20--40$\mu$m)
is slightly over-estimated but a higher spectral resolution
may resolve this excess into the crystalline silicate features.

The differences between our model results and previous work is
in part due to our use of an axisymmetric model,  but also in
part due to the differences in grain parameters.  \citet{skinner94}
used the astronomical silicate \citep{draine84} and 
an MRN grain size distribution \citep{mathis77}. \citet{hrivnak89}
used the silicate opacity function of \citet{volk88} in their
calculations.  \citet{hoogzaad2001} have made the most detailed
analysis of the grain mineralogy to date.  They follow the radiative
transfer of each grain size and type separately allowing a 
more careful analysis of grain composition than our averaged-grain-properties 
approach permits.  

When rescaled appropriately for differences in assumed distance,
our derived physical parameters are comparable to \citet{skinner94}
but significantly different from those of \citet{hrivnak89}.
While our luminosity (L$_*$$\sim$2800 L$_\odot$) and 
$R_{\rm min}$ ($11 \times 10^{15}$cm)
are within 30\% of \citet{skinner94}, our mass-loss rates 
(1.2$\times 10^{-4}$ M$_\odot$ yr$^{-1}$)   are 
a factor of 2 lower than that of \citet{skinner94}.  The reason for
this difference is most likely the difference in dust parameters.
Our grains sizes (0.2--10.0 $\mu$m) 
are significantly larger than those assumed by 
\citet{skinner94} (0.005--1.0 $\mu$m). Larger grains are more efficient 
at emitting far-IR radiation and thus our models require less dust mass
to create the observed far-IR radiation.  The  $R_{\rm min}$, luminosity, and
mass-loss rates, derived by \citet{hrivnak89} are all significantly
larger than our values.  
The predicted $R_{\rm min}$ value of $2.8 \times 10^{16}$cm by \citet{hrivnak89}
is not consistent with the mid-IR image of  \citet{skinner94}.
This large value for $R_{\rm min}$ probably forced the value for L$_*$
higher to create a high enough dust temperature to match the IR
emission.  However, in order to match the optical part of the
SED \citet{hrivnak89} had to include some extinction
due to the ISM  (A$_V$= 0.7), which was not needed by our model
or the model of \citet{skinner94} and is not supported by measurements
of ISM extinction in the direction of this object \citep{neckel80}.
This comparison demonstrates the importance of imaging for independent
constraints of the $R_{\rm min}$ parameter.

\subsection{IRAS~17150$-$3224}

We compare the data and model results for 
IRAS~17150 with  images (Fig.\ref{ir17150image})  and 
SED (Fig. \ref{ir17150sed}).  The best fit model has an
inclination angle of 82\arcdeg~ (see Tables \ref{tabobsprop},
\ref{tabdensfunct}, and \ref{tabdustprop} for other
model parameters). 
In contrast to HD 161796, the central star of IRAS~17150 is 
completely obscured 
by the optically thick dusty torus. Instead we see a spectacular
bipolar reflection nebula in the B band 
image (Fig. \ref{ir17150image}).
The bipolar reflection nebulosity
is the only contributor to the small amount of
optical emission in the  SED (Fig. \ref{ir17150sed}).  
Contours of 9.8~$\mu$m emission show 
the distribution of thermal
radiation from the silicate dust grains and reveal a compact, elliptical
dust nebula associated with the optically obscured region 
(Fig. \ref{ir17150image}; \citealt{meixner99}). The accuracy of
this optical and mid-IR registration is limited by
the absolute position of the mid-IR image  which is known to
about 1\arcsec, however, it is reasonable to assume that the mid-IR
peak is associated with the optically obscuring dusty disk. 
This mid-IR emission traces the inner regions
of the dust shell which extends beyond the optical reflection
nebulosity.  This dust shell, particularly its inner
region, absorbs almost all of the star light and re-emits
it in the infrared.   Hence, the infrared radiation from this dust  
dominates  the spectral energy  distribution of IRAS~17150
 (Fig. \ref{ir17150sed}).

The model images reproduce the spectacular bipolar reflection
nebula and the absence of the central star
(Fig. \ref{ir17150image}). In particular, the size, basic shape and
gap between the lobes are well reproduced, although the
model image intensity is higher in the central region compared
to the data image.  We did not attempt to reproduce the
fascinating arcs observed in this source \citep{kwok98}. The model mid-IR
emission is extended, but slightly smaller than that observed in part
because a Gaussian PSF
does not sufficiently model the wings of the
actual mid-IR psf.  The slight discrepancies between the model and 
data images  may indicate slight differences between the assumed and actual
dust density  distribution.   Our fit to the optical reflection
nebulosity is slightly better than the fit derived by \citet{su00} which
has triangular shaped lobes as opposed to the observed rectangular shape.
The differences in our model images are due mostly to our different 
adopted density distributions.

The images and SED we display here are for an inclination angle
of 82\arcdeg~ which seems to fit the differences in reflection
lobe brightnesses  \citep{su00}. Figure
\ref{ir17150inclang} shows model images for our best fit model
at inclination angles from 0\arcdeg~ to 90\arcdeg. As our view of the nebula
changes from edge-on (90\arcdeg) to pole-on (0\arcdeg), the optical
image transforms from a bipolar nebula with no central star
(60\arcdeg, 90\arcdeg), to a bipolar nebula with a central star 
(30\arcdeg, 45\arcdeg)
to an elliptical or circular nebula with a central star (0\arcdeg--15\arcdeg).
On the other hand, the mid-IR emission changes from a marginally
resolved elliptical nebula which is extended along the bipolar nebula
to an unresolved point source ($\leq$60\arcdeg).  
Comparison of our model images with the data suggest that inclination
angles in the range of 74\arcdeg--90\arcdeg~ could provide an adequate fit and thus
we include an error bar of $\pm$8\arcdeg~ for the
inclination angle.

The model SED compares well with the photometry and spectroscopy data
(Fig. \ref{ir17150sed}), in particular, the silicate absorption feature
at 9.8 $\mu$m is well fit.   
The added ISM extinction of 0.8 A$_V$, which is consistent
with the measurements of \citet{neckel80},
improves the fit to the optical photometry.
The model fits the highest points of the near-IR photometry;
however, we note that there is  a large scatter in these data,
some of which may be due to systematic errors because IRAS~17150 is in 
a crowded field.  The near-IR light is created mostly by scattered
starlight and the excess of this light may indicate that the
dust optical properties, that are derived from lab measurements of
real silicates, produce too much scattering. 
While reasonable, the far-IR fit is slightly high at 
100 $\mu$m and slightly low at 1.3 mm.  
In our model, the sub-mm excess of this source observed
by \cite{hu93}  is created by 
warm, extremely large dust grains  of $\sim$200 $\mu$m in 
radius  in the high density  torus near the central star. 
Figure \ref{ir17150sed}
shows how the spectral energy distribution changes  when one modifies
the number of large grains by decreasing $a_0$ from  200 $\mu$m to
10 $\mu$m while keeping all other
parameters the same.  With $a_0~=~10~\mu$m,
the sub-mm flux is significantly underestimated.
Because we included the large dust grains in our model calculations,
we are able to match the flux of sub-mm photometry points where previous
models underestimated the flux  \citep{su00,hu93}.  Moreover, our
match to the optical photometry is better than previous modeling 
efforts.  \citet{hu93} employed a spherically symmetric model which
does not allow any optical light to escape in a bipolar reflection nebula
and therefore severely underestimates the optical light.  The model SED
of \citet{su00} has a double peaked structure with one peak at 1~$\mu$m
which is quite different than the flattened rolled structure of the
observed SED at the optical wavelengths.  The grain properties adopted
by   \citet{su00} are different from ours which probably explains
most of the difference in our model SEDs.

When corrected for differences in distance,  our model luminosity 
(L$_*$$\sim$27200 L$_\odot$) is
within 10\% of that derived by \citet{hu93}, however the other
parameters are significantly different.  Our inner radius
($R_{\rm min}\sim 9.7\times 10^{15}$ cm) is a factor of two larger than 
\citet{hu93} and our  dust temperature at $R_{\rm min}$ 
($T_{\rm dust}$ = 222 K) is consequentially lower.  
These latter two parameters are well
constrained by the mid-IR image in combination with the SED.
The mass-loss rate derived by \citet{hu93} 
($3 \times 10^{-4}$M$_\odot$ yr$^{-1}$) is comparable to our
derived AGB mass-loss rate  ($5.3 \times 10^{-4}$M$_\odot$ yr$^{-1}$)
but significantly smaller than our derived superwind mass-loss rate
 ($8.5 \times 10^{-3}$M$_\odot$ yr$^{-1}$).  Since our model is
a better fit to the data than \citet{hu93}, the differences
in mass-loss rates reflect the refinement of our model,
especially the axisymmetric geometry, over 
their more simple model.

\section{Discussion}

Comparing the images and SEDs of HD~161796 and IRAS~17150,
one can see that they are quite different.
The derived parameters from the model calculations listed
in Table \ref{tabobsprop} give us some
insight as to why they look different.
The basic geometry used to model both objects  is the same,
the difference lies in the chosen physical parameters for the dust shells.
Interestingly, both  objects  have similar, nearly
edge-on inclination angles.  However, they do differ in 
optical depth of the dust,  mass-loss history and dust grain
composition. 

\subsection{Mass-loss histories}

The apparent differences between HD~161796 and IRAS~17150
can be straightforwardly explained by differences in their
mass-loss histories derived in our model calculations
(see Table \ref{tabobsprop}).
Both objects ended theirs lives on the AGB with
an  axisymmetric superwind.  However, every quantity related
to the mass loss for IRAS~17150 is significantly higher than for
HD~161796.  IRAS~17150's  superwind
mass-loss rate, ${\dot M}_{\rm SW}$, 
 is a factor of 100 larger than HD~161796's while its AGB wind
mass-loss rate, ${\dot M}_{\rm AGB}$, is a factor of
5 higher than  HD~161796's.  For IRAS~17150 there is
a factor of 16 increase in mass-loss rate between the
AGB wind and the superwind but there is no essentially
no average increase for HD~161796 just a change in geometry.  
Note that the superwind mass-loss rates quoted in Table \ref{tabobsprop} 
are latitude averaged,
and that equatorial mass-loss rate is much more intensive than
the polar mass-loss rate.  Thus, the equatorial mass-loss rate
for HD~161796 is much higher during the superwind phase than
during the AGB wind phase.
During the superwind, the equator-to-pole mass-loss ratio
is 160 for IRAS~17150 and only 9 for HD~161796.
While the mass-loss rates for IRAS~17150 were higher than 
for HD~161796, the durations of the
superwind and AGB winds are comparable.
Thus the higher optical depth for IRAS~17150 compared to HD~161796
is caused by the more intensive mass loss experienced by IRAS~17150,
particularly during the superwind phase as it ended its life on the AGB.  
The absolute values for these mass-loss rates depend on the
gas-to-dust mass ratio and the distance 
to the object.  We adopt a gas-to-dust mass ratio of 280  derived
for the OH/IR star  OH 26.5+0.6  by \citet{justtanont96}.  However,
it should be noted that gas-to-dust mass ratios vary by several
factors for O-rich AGB stars  \cite{habing96}.
If IRAS~17150 were actually
closer than 3.6 kpc, then the mass-loss rates would be smaller
than calculated because mass-loss rates depend on the square
of the distance.  However, IRAS~17150 cannot be as close as
HD~161796  because then its luminosity would be smaller
than that of HD~161796 which is inconsistent with IRAS~17150 being
a population I object and HD~161796 being a thick-disk object.
It could also not be too close because then the mass-loss rates
for IRAS~17150 would be  so low  compared to HD~161796 that it is
hard to explain the silicate absorption feature in IRAS~17150.
Hence, despite uncertainties in distances,
the mass-loss rates for  IRAS~17150 must have been higher than
they were for HD~161796.

The total mass for the circumstellar shell of IRAS~17150,
14.0 M$_\odot$, is substantially higher than
the mass for HD~161796, 2.0 M$_\odot$.  This difference in
circumstellar mass points to a difference in progenitor-star mass.
If we assume that the central star masses are $\sim 0.6$M$_\odot$
for HD~161796 and $\sim 1.0$M$_\odot$ for IRAS~17150 \citep{iben83},  then
IRAS~17150's progenitor star was much more massive than HD161796's
in agreement with their population I vs. thick-disk classes.
We note that the total mass for IRAS~17150 (15 M$_\odot$) is higher
than typically expected for an intermediate mass star.  Given the
uncertainties in the gas-to-dust mass ratio and distance for this source,
we do not place great significance to this high number; however, clearly
IRAS~17150 must be on the high mass end of stars that evolve into 
PNs and white dwarfs.

\subsection{Inclination Angle Effects}

Comparison of the Figures \ref{hd161796inclang} and
\ref{ir17150inclang} clearly show that you cannot
view the model for HD~161796 at any inclination angle
and see something like IRAS~17150.  However,  one might
be tempted to think that you could view the IRAS~17150
model pole-on, where the optical depth is low,
and see something like HD~161796.
To test this idea, we have redisplayed the 0\arcdeg~ 
and 15\arcdeg~ inclination angles
for the IRAS~17150 model in Figure  \ref{ir17150map_1kpc} 
to show the appearance
of the structure if the nebula were at a distance of 1 kpc.
These images provide a reasonably accurate comparison with HD~161796
because the $R_{\rm min}$ of the two sources are comparable in our models
and thus the ability to resolve $R_{\rm min}$ would be the same if
the objects were at the same distance.  

The images in Figure \ref{ir17150map_1kpc}
do not look like the data images for HD~161796 (Fig. \ref{hd161796image})
and are a considerably poorer match than our optically thin model
for HD~161796. The 0\arcdeg~ inclination angle (Fig. \ref{ir17150map_1kpc}) 
provides a better
match to the optical data  than the 15\arcdeg~ 
because the central star is centrally located, as
observed; however, the reflection nebula is perfectly round and
has none of the observed ellipticity.  Moreover, the mid-IR emission for
the 0\arcdeg~ image
is perfectly round and does not show the two limb brightened
peaks.   The optical 15\arcdeg~ image (Fig. \ref{ir17150map_1kpc}) 
reveals more emission on the NW side of the star making the central
star appear off-center.  The mid-IR  15\arcdeg~ image 
shows a similar asymmetry as the optical image
with one peak located to the NW of the central star location
at (0,0).  These asymmetries are created by the high optical depths
at optical and 9.8 $\mu$m.  
The peak in the mid-IR arises from the warm dust emission
on the hotter, inner edge of the optically thick dust torus which we
observe through the NW biconal opening that is pointing towards
us.  The optical depth in the mid-IR for this source
is simply too high to see two limb brightened peaks.

Thus the difference
in appearance cannot be attributed to an inclination angle
effect where HD~161796 is viewed pole-on and
IRAS~17150 is viewed edge-on.   Rather,
we can see the central star of HD~161796 whereas
the central star of  IRAS~17150 is invisible because
the dust optical depth along the equatorial plane for HD~161796, A$_V$=1.4, 
is substantially smaller than for  IRAS~17150, A$_V$=37.

This discussion demonstrates the importance of images in differentiating
inclination angle effects and dust shell optical depth effects.
In particular, well resolved mid-IR emission images  can clearly
distinguish between a nearly edge-on optically thin model and a
nearly pole-on optically thick model where just  SEDs alone cannot.
Optically thin, nearly edge-on PPN will show two peaks symmetrically
centered on the star.  Optically thick, nearly pole-on PPN will
show one asymmetric peak or a nearly complete circle.

While inclination angle effects do not explain the differences
between HD~161796 and IRAS~17150,  they are important in modifying
the apparent morphology of PPNs, particularly the DUPLEX PPNs.
\citet{su01} show the important  effects of inclination angle on
the appearance of PPNs that we would classify as DUPLEX PPNs.
Inclination angle effects can cause an ambiguity in classifying
a PPN as  SOLE or  DUPLEX  if the PPN is viewed pole-on.
Comparing the 0\arcdeg~ images of Figs. \ref{hd161796inclang}
and \ref{ir17150inclang}, we find little to distinguish 
the two subclasses in the optical or mid-IR images.

\subsection{Evolution Effects}

Evolutionary effects could also change appearances
because these dust shells are expanding away from the central
star and the optical depth of the dust shell is expected to
decrease over time changing an optically thick object
like IRAS~17150 into one like HD~161796.  We compute dynamical
ages for each source by dividing the $R_{\rm min}$ (Table \ref{tabobsprop}) 
parameter by
the  expansion velocity (v$_{\rm exp}$, Table \ref{tabobsprop}) for each source.
The dynamical ages for both objects, $t_{\rm dyn}$ (Table \ref{tabobsprop}), 
are short but comparable.
HD~161796 left the AGB $\sim$300 years ago.  IRAS~17150 left
210 years ago.  Thus, there is little time for an object
like IRAS~17150 to evolve into an object like HD~161796.

\subsection{Dust Composition}

One of the unanticipated results of our modeling  was
the significant difference in dust composition between
HD~161796 and IRAS~17150.
HD~161796 has mostly amorphous silicates,
but also crystalline silicates and crystalline water ice.
In contrast, IRAS~17150 seems adequately fit with just amorphous silicate,
and recent ISO spectroscopy presented by \citet{volk01} shows no clear
evidence for crystalline silicates.
The grain sizes are remarkably different. IRAS~17150 
has a wide range of grain sizes with the minimum size of 
0.001 $\mu$m and the effective maximum size of 200 $\mu$m.
HD~161796 has a narrower range of grain sizes with the minimum size of
0.2 $\mu$m and the effective maximum size of 10.0 $\mu$m.  
During our modelling efforts, we  started with a single grain size
which proved to be inadequate, in particular for IRAS~17150.  We also
varied the $a_{\rm min}$ and $a_0$ values and found them to be quite
sensitive ($\sim$10\% level) to the mid-IR and far-IR ends of the SED, 
respectively.   While the exact values for $a_{\rm min}$ and $a_0$ are 
uncertain due to the inexactness of the dust composition, it is unlikely that
this uncertainty is large enough to make the possible size distributions for
these two objects the same.

The extremely large grain
sizes found for IRAS~17150 are similar to that
found for the Red Rectangle, another DUPLEX object,
by \cite{jura97}.  Large grains were hypothesized to form
in the long-lived disk of the Red Rectangle to explain
their presence.  It may be that IRAS~17150 also
has a long lived circumstellar disk.  Our model certainly
reveals an extreme equatorial density enhancement which
is essentially a disk.  However, it is not clear if this disk
is gravitationally bound to IRAS~17150 as it is with
the Red Rectangle and kinematic evidence is needed to establish
such a fact.

\citet{molster99}
and \citet{molster00} have shown that PPN with disks have stronger
crystalline silicate band strengths than  PPN without disks which 
they classify as  outflow sources.  In their studies,
HD~161796 is considered an outflow source which is independently
supported by our model calculations.  IRAS~17150 was not included
in the study by \citet{molster99}
and \citet{molster00}.  HD~161796 has crystalline silicates while IRAS~17150
does not.  Possibly, IRAS~17150 is early in its evolution
and may develop prominent crystalline silicate bands as its central star
evolves to hotter temperatures and the disk has time to evolve.  
In this scenario, large dust grains would evolve in the disk 
before the crystalline silicates.   Alternatively, large grains may simply
form when mass loss rates are extraordinarily high because 
the higher densities may create environments conducive
for growing large grains.   In this scenario,  crystalline silicates
form as the disk evolves.  It is also important to keep in mind
that absence of evidence is not evidence of absence.  \citet{kemper01}
has shown that temperature differences between the amorphous
and crystalline silicates can hide crystalline silicates in warm shells.
In our model of IRAS~17150, the optical depth  at 25 $\mu$m is $\sim$7
and its possible that this high optical depth veils the crystalline
silicate features.

\subsection{Two Subclasses of PPN}

If our model calculations reflect a more general truth about 
PPNs,  then  there is a real physical difference between the
DUPLEX and SOLE  classes of PPN suggested by \citet{ueta00}.    
DUPLEX PPNs, such as IRAS~17150,   originate from more massive
progenitor stars than SOLE PPNs, such as HD~161796.  DUPLEX PPNs
experience more intensive and equatorially enhanced  superwinds  
than SOLE PPNs.   Detailed model calculations of several other
PPNs by us and others \citep{meixner97,skinner97,su98,dayal98,ueta01b}
show similar general differences between SOLE and DUPLEX PPNs,
as classified by \citet{ueta00}. However, future detailed modeling of
other PPN may well reveal PPN with physical parameters that
fill-in a spectrum where IRAS~17150 and HD~161796 mark
the extreme end points.

We  might expect there to be  subclasses of
PPN because there are subclasses of PN morphologies , e.g. bipolar,
elliptical, and round \citep{balick87,manchado96}.  
Bipolar PN originate from higher mass progenitors than elliptical
PN  as suggested by their differences in galactic scale height
\citep{corradi95}.
Higher mass star progenitors of PNs are expected to
lose mass at higher rates  because
main sequence stars initially of 0.8 to 8 solar masses 
become white dwarfs which
have masses narrowly peaked at 0.6 solar masses \citep{iben83}.
Our model calculations presented here support the suggestion  of \citet{ueta00}
that DUPLEX  and SOLE PPNs are 
precursors to bipolar and elliptical PNs, respectively.  

The differences in mass-loss histories between DUPLEX PPNs (bipolar PN)
and  SOLE PPNs (elliptical PN) may also suggest that there is a difference
in mass-loss mechanisms.   Progenitor AGB stars of
DUPLEX PPNs  have a much more efficient means
to loose their  outer envelopes than SOLE PPN progenitors.  
It also appears that
mass-loss is more efficient when it is axisymmetric.   Whether this
mass-loss enhancing process is  interaction with a binary star companion
or magnetic fields,  a sudden increase in mass-loss rate coincident
with an axisymmetric geometry must be explained.

\section{Conclusions}

Our radiative transfer models of
HD~161796, a SOLE PPN,  and IRAS~17150, a DUPLEX PPN,
show that physical differences, not  inclination angles,
create their observed differences.  HD~161796 and IRAS~17150
are both viewed nearly edge-on. However, IRAS~17150  experienced 
mass loss with a higher rate 
and with a higher equator-to-pole mass-loss ratio than HD~161796.
This difference in mass-loss  history created a larger
optical depth of dust in IRAS~17150 than in HD~161796.  
If these calculations reflect a more general truth about PPNs, then they
support the hypothesis put forth by  \cite{ueta00} that SOLE PPN and
DUPLEX PPN are physically distinct subclasses of PPNs.
SOLE PPNs are probably descended from lower mass progenitor
stars and are the precursors to elliptical PNs.
DUPLEX PPNs are probably descended from higher mass  progenitor
stars and are the precursors to bipolar PNs.
The dust compositions of both IRAS~17150 and HD~161796 are dominated by
large amorphous silicate grains; however, the grain sizes have a much
larger range  for IRAS~17150 (0.001$\mu$m $<$ a $\ale$~200 $\mu$m) 
and  HD~161796 contains a
significant component of crystalline silicates and water ice.

\acknowledgements
We are grateful for the insightful comments of the referee who
gave the paper a very thorough review and to M. Jura.
Meixner,  Ueta and Speck are supported by NASA Grants GO-06737.01-95A,
GO-06364.02-95A and STI 7898.02-96A and NSF Career Award Grant AST 97-33697.
Bobrowsky is supported by STScI Grants GO-06364.01-94A and
GO-006737.02-95A.


\begin{deluxetable}{ccc}
\scriptsize
\tablewidth{40pc}
\tablecaption{\label{tabsoledup} Observed differences between SOLE and DUPLEX PPNs}
\tablehead{
\colhead{Property} &  \colhead{SOLE} & \colhead{DUPLEX}
}
\startdata
optical morphology & star + nebula  & nebula dominant \nl
mid-IR morphology & toroidal & core-elliptical \nl
SED &  optical$\sim$ far-IR & far-IR dominant \nl
Galactic height & $\ale$ 2100 pc & $\ale$ 520 pc \nl
\enddata

\end{deluxetable}

\begin{deluxetable}{cccc}
\scriptsize
\tablewidth{40pc}
\tablecaption{\label{tabobsprop} Properties of the two selected PPN from observations and best-fit models}
\tablehead{
\colhead{Quantity} &  \colhead{HD 161796} & \colhead{IRAS~17150$-$3224}
& \colhead{Ref.}}
\startdata
$\rm L_{*} \propto d^{2}$ $(L_\odot)$     & 2800  & 27200    \\
T$_*$ (K) ($\pm$10\%)    & 7000 & 5200 &  1,2 \nl
$\rm R_{*} \propto d^{2}$ ($R_\odot$)  & 34 & 201 \\
d (kpc)  & 1.0 & 3.6 & 3,4 \nl
ISM A$\rm _V$ & 0.0 & 0.8  & 5 \nl
$\rm T_{dust}$ at R$_{\rm min}$ (K) & 110 & 220  \\
$R_{\rm min} \propto d$ (10$^{15}$ cm) & 11 &  9.7  \\
v$_{\rm exp}$ (km/s)  & 12.0 &  15.0  & 6,2 \nl
{\it i} & $90^\circ \pm 30^\circ$ & $82\pm 8^\circ$ \\
A$_{\rm V}$ at pole    & 0.80 &  0.98  \\
A$_{\rm V}$ at equator & 1.4 & 37 \\
Shell Mass $\rm \propto {\rm d}^2$  (M$_\odot$) & 2.0  &  14.0 & 7 \\
${\dot M}_{\rm AGB} \propto {\rm d}^2$ (10$^{-4}$M$_\odot$
yr$^{-1}$) & 1.2 & 5.3  & 7  \\
t$_{\rm AGB} \propto$ d ($10^4$ yrs) & 1.7 & 2.0  \\
${\dot M}_{\rm SW} \propto d^2$ (10$^{-4}$M$_\odot$ yr$^{-1}$) & 1.2
& 85  & 7 \\
t$_{\rm SW} \propto$ d (yrs) & 590 & 410 \\
t$_{\rm dyn} \propto$ d (yrs)  & 300 & 210  \\
${\dot M}_{\rm equator} / {\dot M}_{\rm pole}$ & 9  & 160  \\

\enddata

\tablerefs{ 
$^1$\cite{hrivnak89},
$^2$\cite{hu93},
$^3$ \cite{skinner94},
$^4$ this work,
$^5$\cite{neckel80},
$^6$\cite{likkel87},
$^7$ Assumed gas-to-dust mass ratio of 280 \cite{justtanont96}.
}
\end{deluxetable}

\begin{deluxetable}{ccc}
\scriptsize
\tablewidth{40pc}
\tablecaption{\label{tabdensfunct} Parameter values for the density functions
of best-fit models }
\tablehead{
\colhead{Parameter} &  \colhead{HD~161796} & \colhead{IRAS~17150$-$3224}
}
\startdata
$A$ & 8 & 159 \\
$B$ & 2 & 2 \\
$C$ & 2 & 1.5 \\
$D$ & 1 & 1 \\
$E$ & 3 & 4 \\
$F$ & 1 & 1.5 \\
$\rm \tau_{9.8\micron}$ at equator & 0.46 & 12.0  \\
$\rm \tau_{9.8\micron}$ at pole    & 0.26  & 0.32  \\
\enddata
\end{deluxetable}

\begin{deluxetable}{cccc}
\scriptsize
\tablewidth{40pc}
\tablecaption{\label{tabdustprop} Dust  Properties for best-fit models}
\tablehead{
\colhead{Quantity} &  \colhead{HD~161796} & \colhead{IRAS
17150$-$3224}& \colhead{Ref.}
}
\startdata
$a_{min}$ ($\mu$m)  &  ~0.2    &  ~~~0.001 \\
$a_{0}$ ($\mu$m) &  10.0 &  200.0 \\
amorphous silicate  & 86.5\%   &  100\% & 1 \\
crystalline enstatite   & 6\%   &  0\% & 2\\
crystalline  forsterite  & 5\%  & 0\% & 3 \\
 water ice      & 2.5\%  & 0\% & 4\\

\enddata

\tablerefs{$^1$\cite{dorschner95},
$^2$crystalline enstatite \cite{scott96,jaeger98},
$^3$crystalline forsterite \cite{scott96,servoin73}, 
$^4$water ice \cite{warren84,bertie69}.} 
\end{deluxetable}

\newpage

{\bf Figure Captions:}

\figcaption[f1.eps]{Grayscale schematics of the PPN density function used
in the model calculations. The grayscale wedges are in units
of g cm$^{-3}$. The central diagram shows
the locations of the axisymmetric superwind and 
spherically symmetric AGB wind with $R_{\rm SW}$ as the
boundary.  The inner radius, $R_{\rm min}$, and the outer
radius, $R_{\rm out}$, define the inner and outer boundaries
of this density function.  The left image shows the density
function used in the model of HD~161796.  The right image shows
the density function used in the model of IRAS~17150.\label{densfunct}}

\figcaption[f2.eps]{Data (left) and model (right) images for HD~161796. 
The contours show the thermal dust emission at $12.5~\mu$m. 
Contour levels are 10, 30, 50, 70, and 90\% (82.5 and 90\%
for data) of the peak. 
Peak values are 0.95 Jy arcsec$^{-2}$ for the data and 
2.08 Jy arcsec$^{-2}$ for the  model. 
The grayscale image shows the scattered and direct star light 
at V-band ($0.55~\mu$m) displayed in log-scaled intensity 
units of mJy pixel$^{-1}$ where the pixel scale is 
0$\arcsec$.023 and 0$\arcsec$.046 for data and model, respectively.
The HST V-band image, which shows the diffraction spikes in its
psf,  is from  \cite{ueta00} and the $12.5~\mu$m image is from 
\cite{skinner94}.\label{hd161796image}}

\figcaption[f3.eps]{The spectral energy distribution of HD~161796.  
The model calculated SED is shown in dashed lines. 
The ISO mid-IR spectra (solid line) is from 
\cite{molster00} and \cite{hoogzaad2001}. 
The photometry data (squares) are from a compilation 
by \cite{ueta00}, which include data from 
\cite{humphreys74,fernie83,hrivnak89,skinner94,meixner99,ueta00}
and IRAS.
The inset shows the IR peak around $40~\mu$m. \label{hd161796sed}}

\figcaption[f4.eps]{Model  images for HD~161796 shown at inclination
angles 0 to 90\arcdeg.  
The contours show the thermal dust emission at $12.5~\mu$m
and the grayscale shows the optical V-band images. \label{hd161796inclang}}

\figcaption[f5.eps]{Data (left) and model (right) images for 
IRAS~17150$-$3224. 
The contours show the thermal dust emission at $9.8~\mu$m. 
Contour levels are 20, 30, 50, 70, and 90\% of the peak. 
Peak values are 27 Jy arcsec$^{-2}$ for the  data 
and 39.3 Jy arcsec$^{-2}$ for the model. 
The grayscale images show the reflection nebula at B-band 
(0.45 $\mu$m) displayed in log-scaled intensity units of 
mJy pixel$^{-1}$ where the pixel scale is 
0$\arcsec$.023 and 0$\arcsec$.046 for data and model, respectively.
The B-band HST image is from \cite{ueta00}
and the $9.8~\mu$m is from \cite{meixner99}.\label{ir17150image}}

\figcaption[f6.eps]{The spectral energy distribution of IRAS~17150$-$3224.
The model calculated SED is shown in dashed lines ($a_0 = 200~\mu$m). 
The mid-IR spectra are 
from \citet{kwok95} (solid lines) 
from the IRAS LRS (solid lines). 
The photometry data (squares) are from the compilation of 
several sources by \citet{ueta00}, which include
data from \citet{veen89,hu93,kwok96,reddy96,meixner99,ueta00}
and IRAS.
The inset shows a closeup of the $9.8~\mu$m silicate 
absorption feature.
The dot-dashed line represents another model with $a_0 = 10~\mu$m,
which shows the  dependence of far-IR and sub-millimeter
continuum emission on larger grain sizes. \label{ir17150sed}}

\figcaption[f7.eps]{Model  images for  IRAS~17150$-$3224 shown at inclination
angles 0 to 90\arcdeg.  
The contours show the thermal dust emission at $9.8 \mu$m
and the grayscale shows the optical B-band images.\label{ir17150inclang}}

\figcaption[f8.eps]{Model images for IRAS~17150$-$3224 rescaled for a distance
of 1 kpc for comparison with HD~161796 images.  Inclination
angles of 0 and 15\arcdeg~ shown. The contours show the thermal 
dust emission at $9.8~\mu$m
and the grayscale shows the optical B-band images.\label{ir17150map_1kpc}}

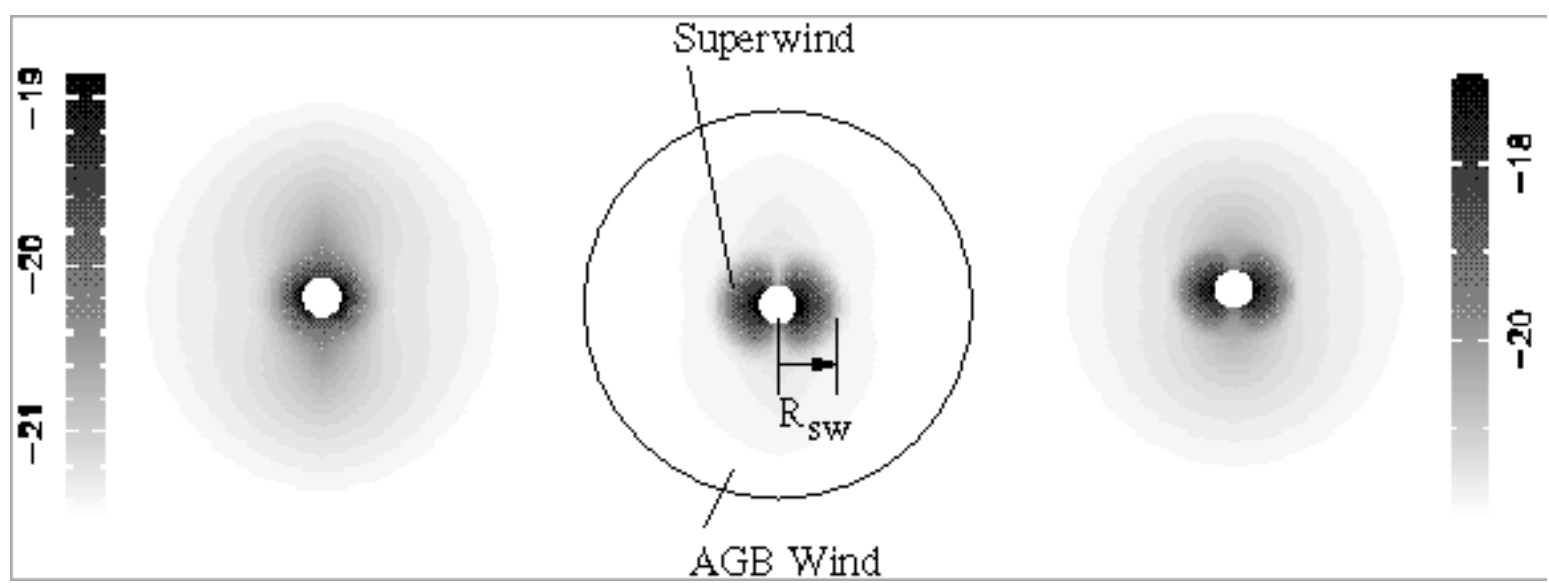
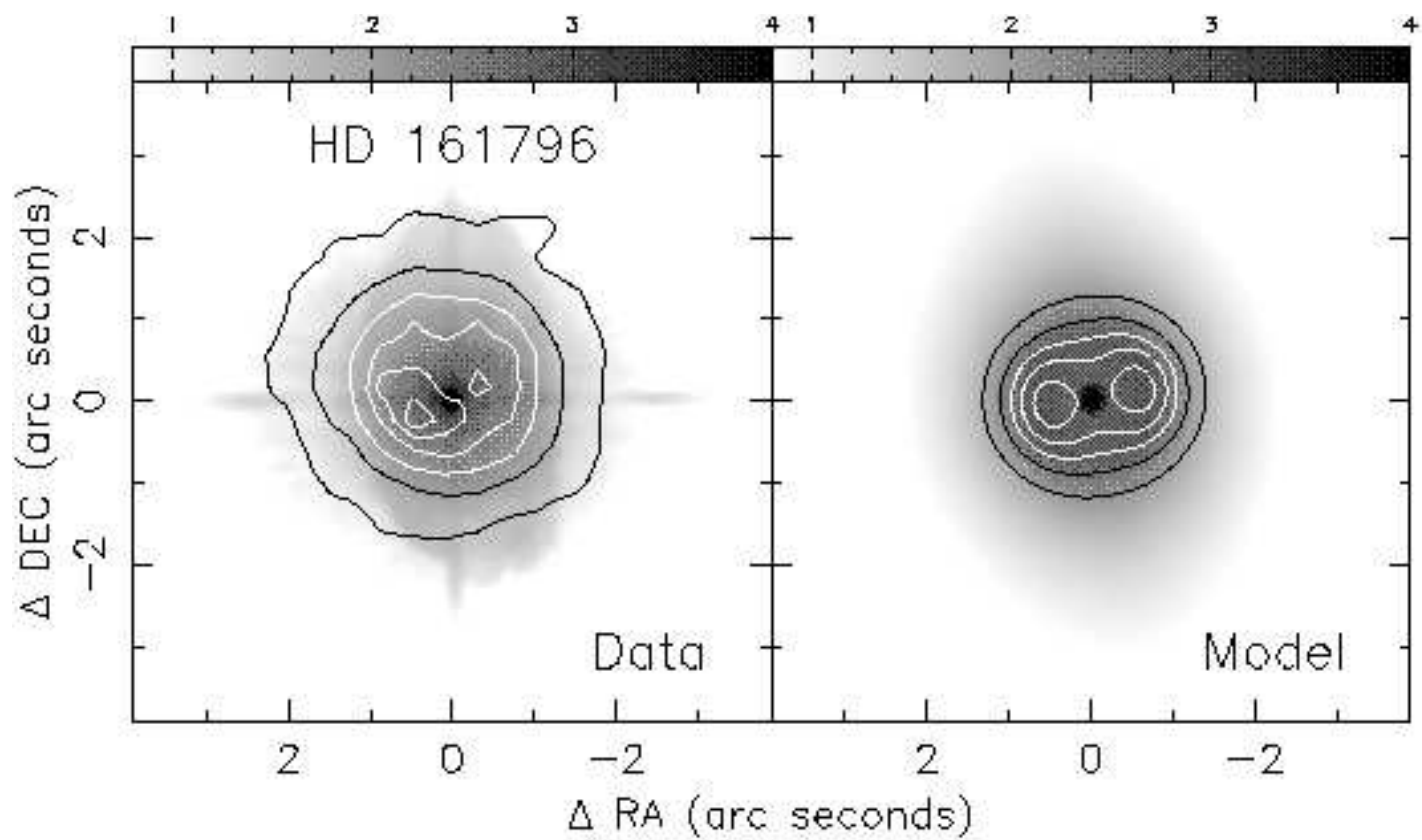
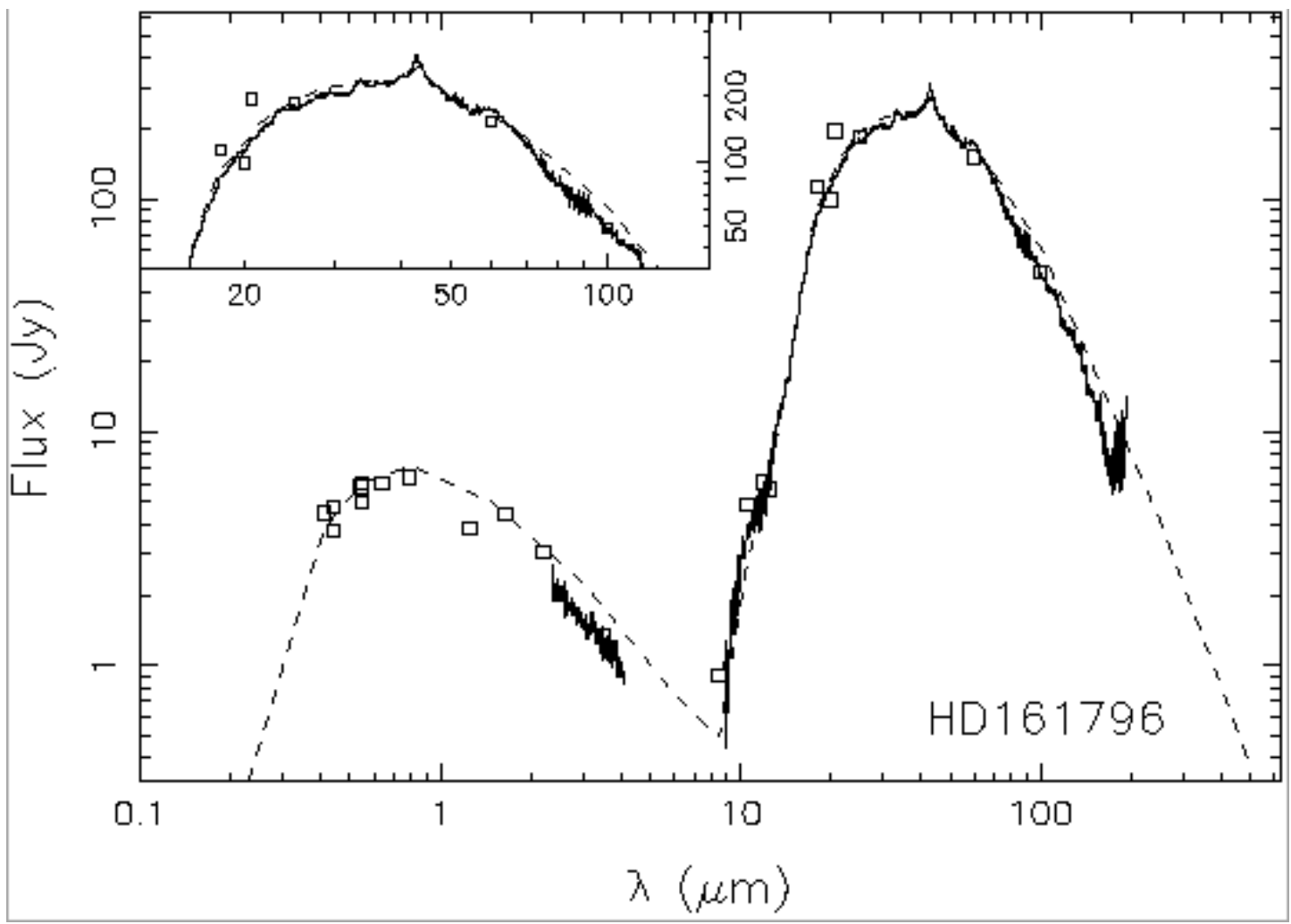
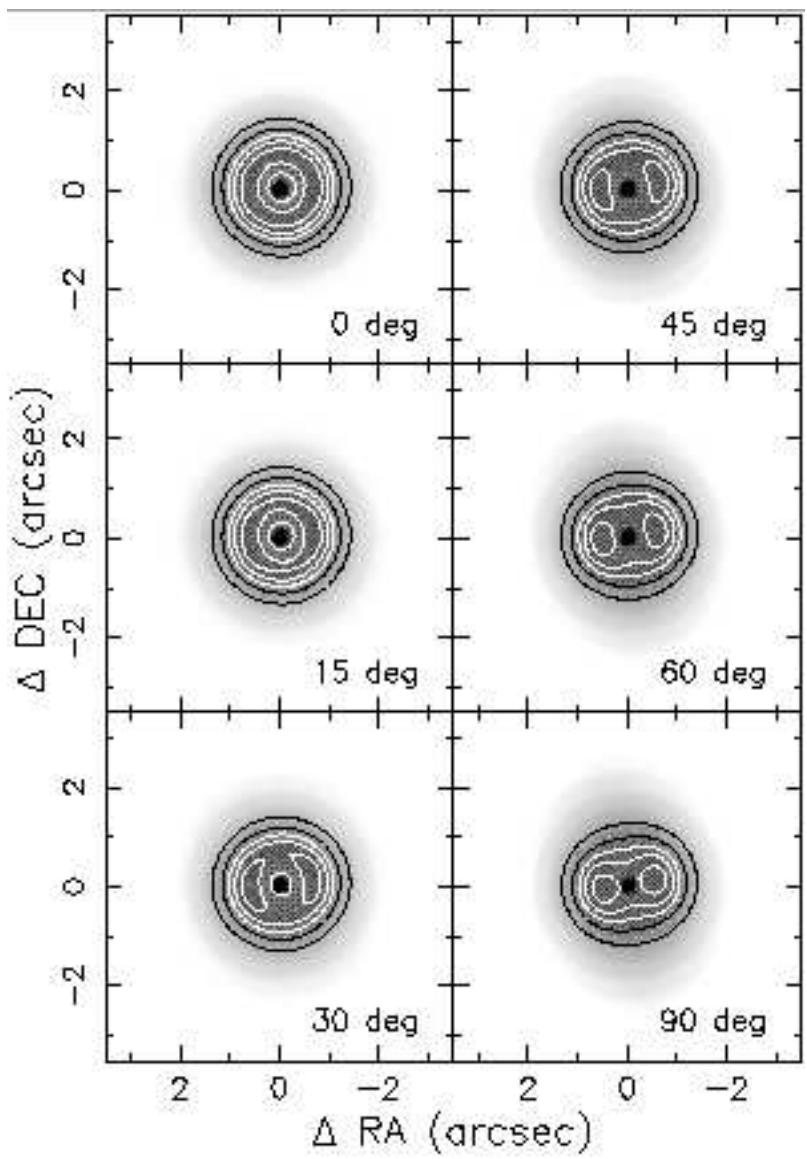
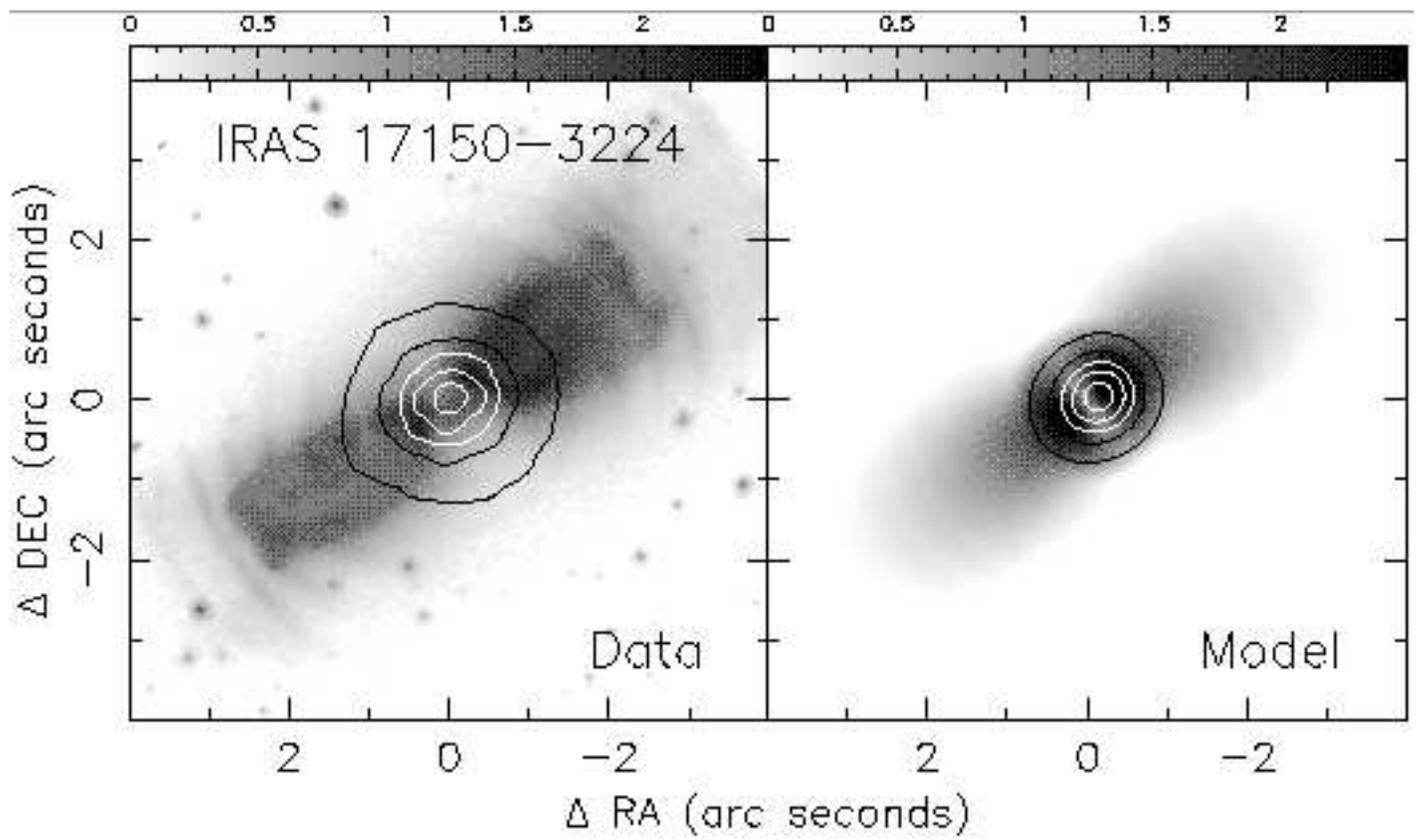
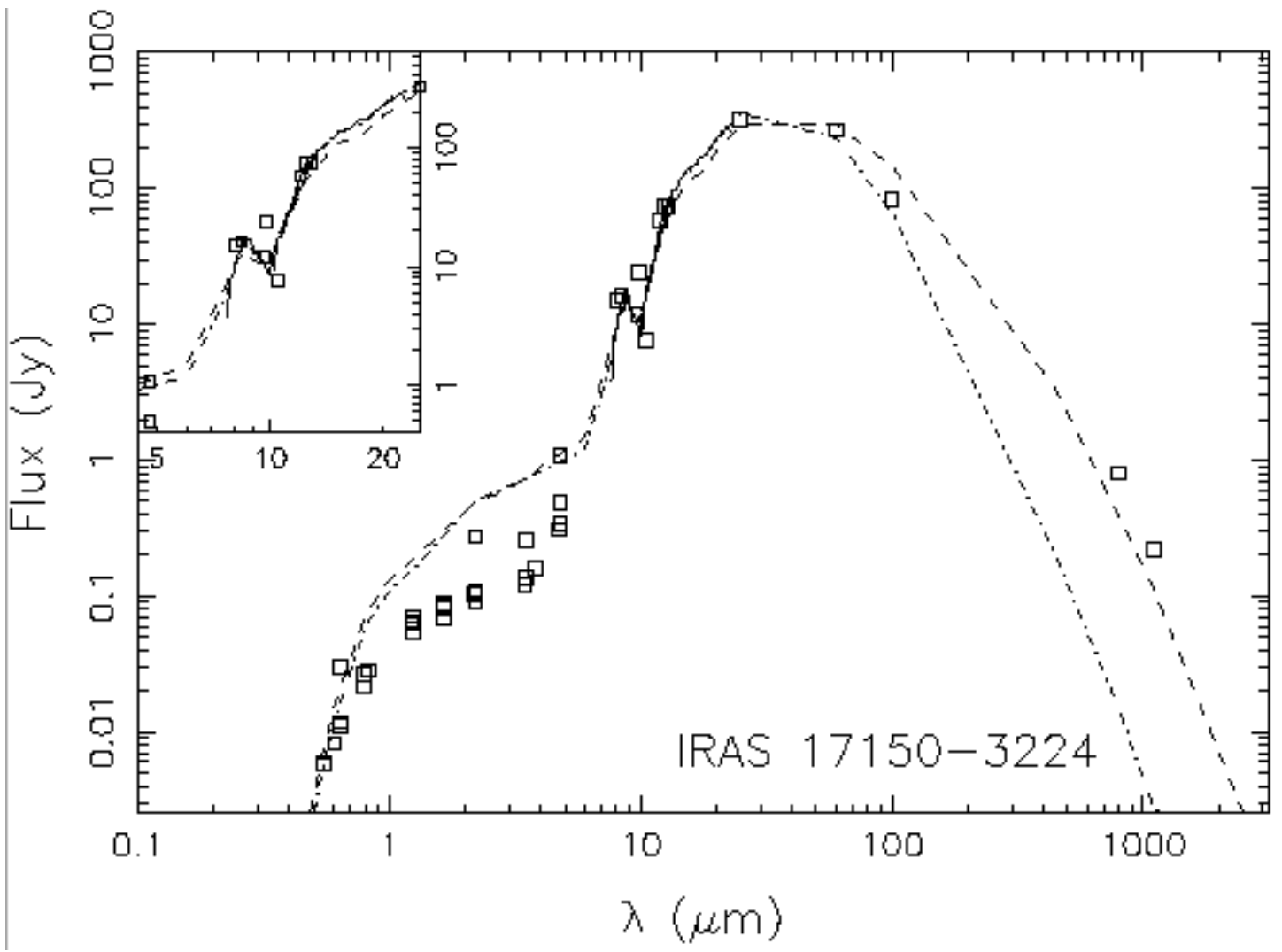
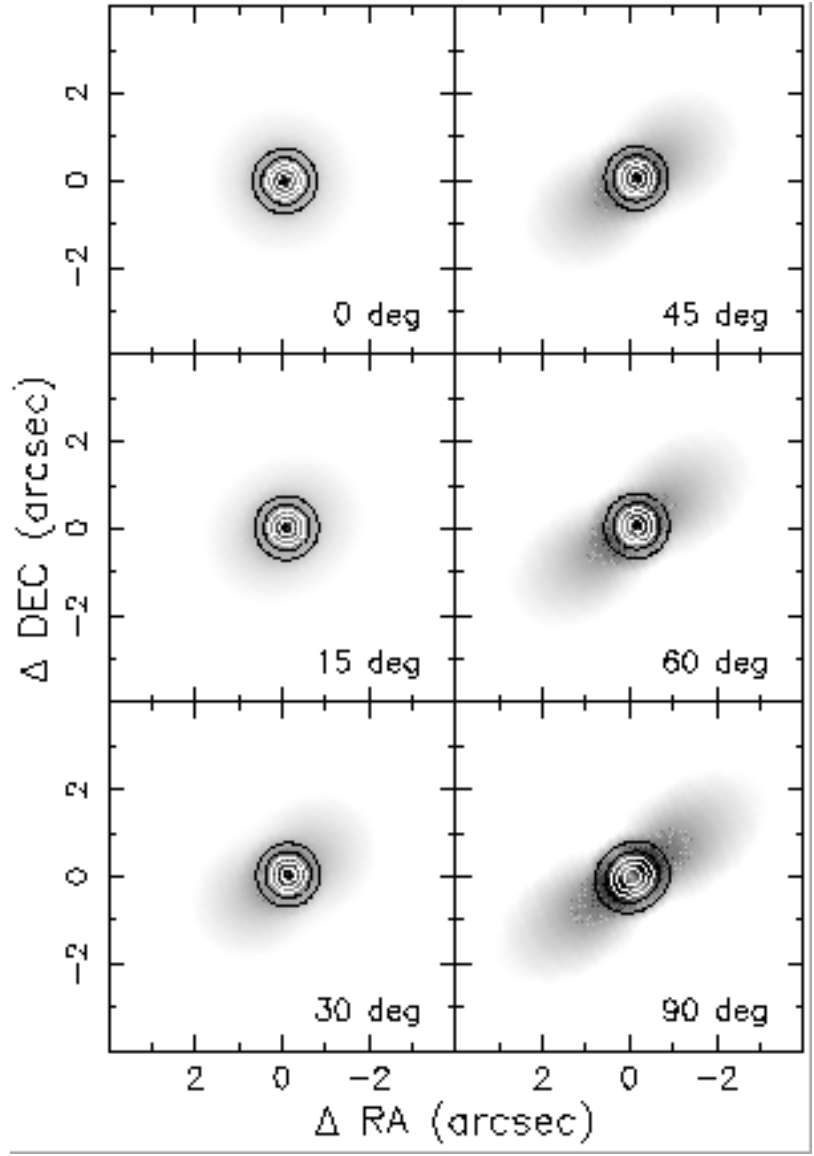
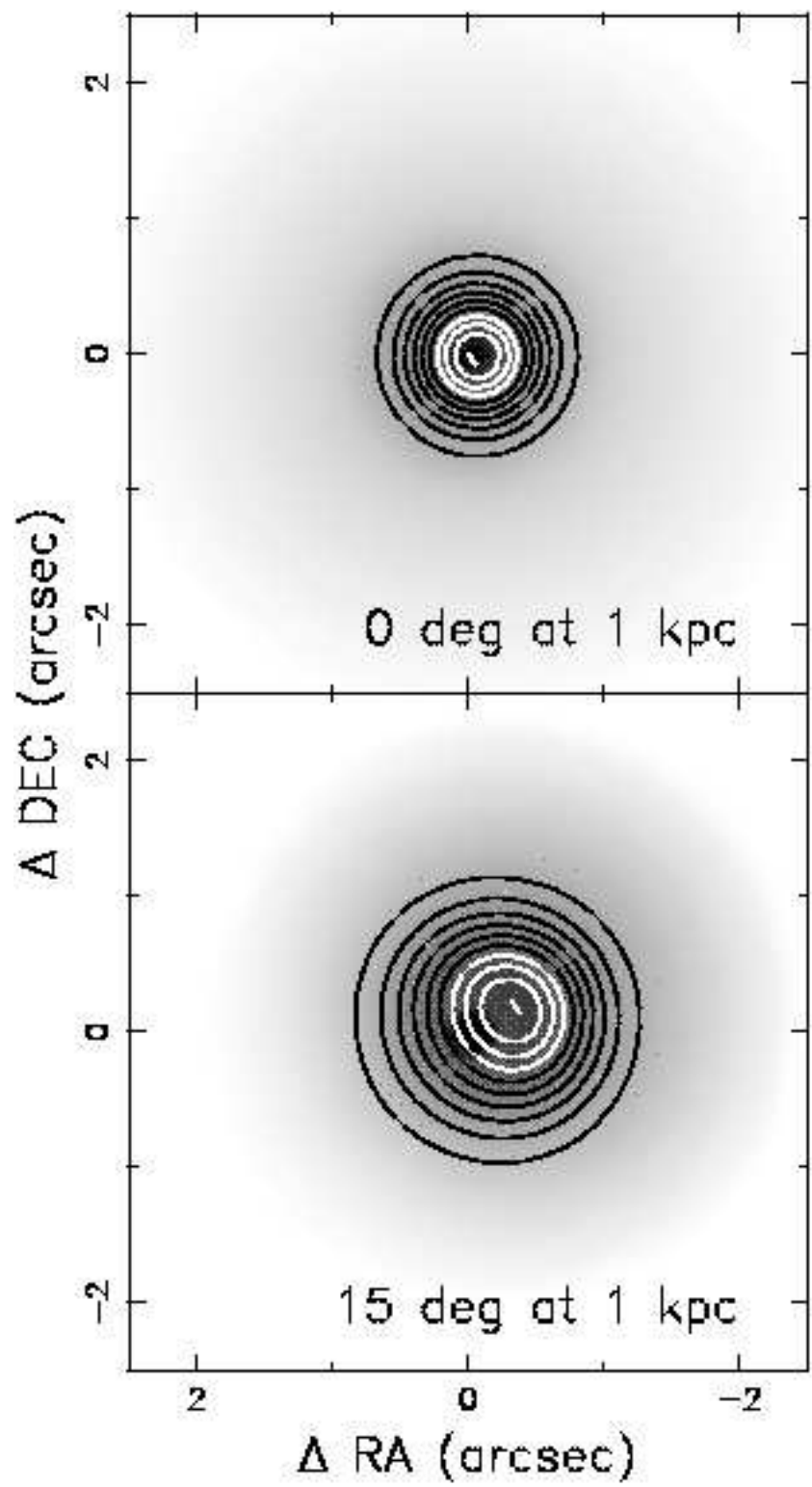
\end{document}